# Microscopic structure differences in CZTSe quaternary alloys prepared by different techniques revealed by spatially-resolved laser-induced-modification Raman spectroscopy

Qiong Chen[1], Sergio Bernardi[2], and Yong Zhang[1, †]

[1]Department of Electrical and Computer Engineering, and Energy Production and Infrastructure Center (EPIC), The University of North Carolina at Charlotte, Charlotte, NC 28223, USA

[2]Semiconductor Materials Specialist, C.so Trapani 10, 10139 Turin, Italy.

While producing comparable efficiencies and showing similar properties when probed by conventional techniques, such as Raman, photoluminescence and X-ray diffraction, two thin film solar cell materials with complex structures, such as quaternary compound CZTSe, may in fact differ significantly in their microscopic structures. In this work, laser induced modification Raman spectroscopy, coupled with high spatial resolution and high temperature capability, is demonstrated as an effective tool to obtain important structure information beyond that the conventional characterization techniques can offer, and thus to reveal the microscopic scale variations between nominally similar alloys. Specifically, CZTSe films prepared by sputtering and co-evaporation methods that exhibited similar Raman and XRD features were found to behave very differently under high laser power and high temperature Raman probe, because the differences in their microscopic structures lead to different structure modifications in response to the external stimuli, such as light illumination and temperature. They were also shown to undergo different degree of plastic changes and have different thermal conductivities as revealed by spatially-resolved Raman spectroscopy.

[†] Corresponding author: yong.zhang@uncc.edu

# I. INTRODUCTION

Raman spectroscopy is an effective tool for chemical composition and structural analyses of materials, assuming that the probe light intensity is sufficiently weak such that no structural change has been induced by the perturbation. While Raman spectroscopy probes the material structure through revealing the vibrational finger prints of the material, transmission electron microscopy (TEM) is generally considered to be a more direct probe for the material structure. However, the probing volume of TEM tends to be small, typically in the order of a few tens of nanometer, which places a severe limit on the characterization efficiency when we need to have a macroscopic scale structural information of an inhomogeneous material. Two extreme versions of laser spectroscopy techniques have been developed for a larger volume structural analysis. One is called laser induced breakdown spectroscopy (LIBS) [1], which uses a highly energetic laser pulse as the excitation source to excite and atomize the sample, then measures the light emission of the atoms to determine the chemical composition of the material. In the process, the material is ablated at the excitation site. The other one is called femtosecond laser tomography [2, 3], which also uses a highly energetic ultra-fast laser pulse to assist the evaporation of atoms, then time-of-flight mass spectrometry equipped with position sensitive detectors determine the element numbers of the analyzed atoms and their spatial distribution in the sample. These two laser-based material analysis techniques are destructive and do not offer the chemical bonding information when the atoms were in the material. In contrast, the "high power (HP)" Raman spectroscopy, as a special form of laser induced modification spectroscopy (LIMS), uses a tightly focused CW laser with a power density that is just high enough to induce a local structural change but usually without major material ablation, and measures the change in Raman features, compared to the spectrum before the illumination, under "low power (LP)". Performing spatially resolved Raman mapping on an as-grown material can already reveal composition and/or structural variations in the sample. However, the HP Raman spectroscopy can provide additional information beyond that obtainable from the conventional technique. For instance, some structural or chemical fluctuations are too feeble to be detectable or distinguishable in the as-grown sample or between different ones, but are signified after being modified by the HP illumination. The appropriate power levels of "HP" and "LP" depends rather sensitively on the specific material of interest [4]. The application of the HP Raman spectroscopy can reveal some subtle but important structural differences in two samples that might otherwise appear to be indistinguishable if they were only subjected to the conventional probes, such as Raman, PL, XRD, and device characterization.

Additional or complementary information can be obtained by further applying "high temperature (HT)" Raman spectroscopy. By "high temperature" we mean that the temperature is high enough to induce some structural changes associated with the hidden structural/composition fluctuations. The combination of the two approaches or HPHT Raman spectroscopy, in conjunction with the "high spatial resolution (HR)" of the confocal diffraction limit detection, we have a very powerful "3H" laser spectroscopy approach for probing the structural variation and inhomogeneity, in particular for a complex alloy like $Cu_2ZnSnSe_4$ (CZTSe), with much greater sensitivity and efficiency.

The emerging PV material CZTSe may be grown by various different techniques but none of them has yet been able to achieve an efficiency close to Shockley-Quisser limit [5]. This approach offers another useful tool for assessing the potential of an individual technology path by providing an efficient way to distinguish the materials that are otherwise indistinguishable when the conventional characterization tools are applied.

## II. EXPERIMENTAL DETAILS

All Raman measurements were performed with a Horiba Jobin Yvon HR800 confocal Raman system using a 532 nm laser. The spectral dispersion is 0.44 cm$^{-1}$/pixel with a 1200g/mm grating used. The room temperature measurement was carried out with a 100x objective lens with a numerical aperture NA = 0.9. The diffraction limit laser spot size is 0.76 µm, and the spatial resolution of the measurement is about half of the laser spot size. A 50x long working distance lens, with NA = 0.5, was used in high temperature Raman measurements. The laser power was adjusted by either using neutral density filters or charging laser current. For HT measurement, samples were heated by Linkam TS1500 heating system with temperature control accuracy of 1 °C. The HP illuminated site was examined by Scanning Electron Microscopy (SEM) as well as Energy Dispersive X-ray Spectroscopy (EDS) elementary analysis.

Three different CZTSe sample were measured: bare CZTSe film CZTSe_97 (S1); bare CZTSe film M3599_12 (S2), and CZTSe solar cell device M3602_21 (S3). For S1, the CZTSe film was prepared by selenizing the metal stacks which composed of 300 nm Mo, 130 nm Zn, 155 nm Cu and 180 nm Sn sputtered on an Ashai PV 200 glass substrate. The film compositions were [Zn]/[Sn] $\cong$ 1.25 and [Cu]/([Zn]+[Sn]) $\cong$ 0.85 in kesterite phase. The film was slightly thicker than 1 µm. A device with a nominally similar absorber was measured with ~5% efficiency. S2 and S3 films were grown in vacuum with a co-evaporation method on a soda lime glass substrate, with 1 µm sputtered Mo back contact and 150 Å e-beam evaporated NaF precursor. The film thickness was 1.49 µm with [Zn]/[Sn] $\cong$ 1.32 and [Cu]/([Zn]+[Sn]) $\cong$ 0.82 for S2, and 1.3 µm with [Zn]/[Sn] $\cong$ 1.28 and [Cu]/([Zn]+[Sn] $\cong$ 0.87 for S3. The device sample S3 was finished with a chemical bath deposited CdS layer, a sputtered resistive/conductive ZnO bi-layer, e-beam evaporated Ni/Al grids and an MgF$_2$ antireflective coating. Details for the film growth and device processing can be found in Refs. [6] and [7]. A device with nominally same absorber of S2 had an energy conversion efficiency of ∼8% and the efficiency of S3 was also ∼8%.

## III. RESULTS AND DISCUSSIONS

### A. Raman studies at room temperature

#### 1. High-power effects at individual locations

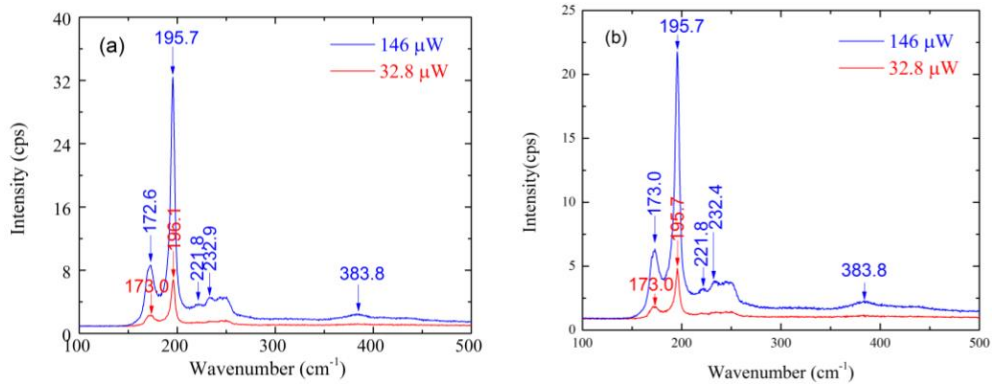

FIG. 1. Raman spectra of the bare CZTSe films measured at 32.8 µW and 146 µW: (a) S1; (b) S2.

In Fig. 1, Raman spectra from S1 and S2 front surface were obtained with both 32.8 µW (7.9 x $10^3$ W/cm$^2$) and 146 µW (3.5 x $10^4$ W/cm$^2$), from the same sample point with no grating movement between the two measurements. For S1, the primary CZTSe main Raman modes (196.1 and 173.0 cm$^{-1}$) experienced a 0.4 cm$^{-1}$ red shift when laser power was increased from 32.8 µW to 146 µW. The observed red-shift was reversible or the change was elastic, when re-measured with the lower power. However, for S2, no shift was observed with the power change. The slight CZTSe Raman mode redshifts indicate that a small heating effect was introduced for the sputtered sample under 146 µW illumination, but the power level was not high enough to cause irreversible material modification. The comparison suggests some subtle structural difference between the two samples, manifested in the difference in thermal conductivity. Despite of the small heating effect found in S1, we will still consider 146 µW as a LP excitation level. Note that within each sample, the absolute position of the Raman mode may vary up to about 0.5 cm$^{-1}$ due to sample inhomogeneity.

Fig. 2 examines the effects of HP illumination by comparing the LP Raman spectra from the same location at 0.146 mW before and after 100s illumination at 2.47 mW (5.9 x 10$^5$ W/cm$^2$), and further 36 second at 4.5 mW (1.1 x 10$^6$ W/cm$^2$). After the 2.47 mW illumination, the illuminated spot showed some color change under optical microscope, but no apparent ablation. However, the 4.5 mW illumination typically resulted in some local material ablation. We note that for a bulk Si even illuminated with the full power (~20 mW) there was practically no change when returned to the LP condition.

The red curves in Figs. 2(a)-(c) are typical Raman spectra respectively from the three CZTSe samples measured firstly at 0.146 mW, and the green and blue curves are the corresponding spectra re-measured after the two higher power illuminations, respectively. In their initial states, all three samples in fact exhibited rather similar Raman features: two intense Raman peaks at 195 – 196 cm$^{-1}$ and 172 – 173 cm$^{-1}$, and a weak peak at 232 – 233 cm$^{-1}$, which are very close to the Raman peaks at 196, 173 and 231 cm$^{-1}$ reported for single crystal CZTSe [8]; additional weak peak appear at ~222, ~245 and ~251.5 cm$^{-1}$. In Fig. 2(d), the spectral region of the multiple weak peaks ("mesa-like" band to the right of the 195 – 196 cm$^{-1}$ peak) is compared directly between the three samples in their initial states, indicating that there was no significant or distinct difference between the samples prepared differently. As a matter of fact, more variations could be found within one sample due to inhomogeneity, as shown in Fig. 2(e) comparing multiple locations of S2. Therefore, it is impractical for the conventional Raman probe to reliably reveal potential structural variations from sample to sample. The ~222 cm$^{-1}$ peak is believed to be a transverse optical (TO) Raman mode in E symmetry from kesterite structure based on theoretical calculations [9-11]. The ~245 cm$^{-1}$ peak was also reported to be CZTSe related both experimentally and theoretically [12-14]. The ~251.5 cm$^{-1}$ peak is close to the Raman modes of these secondary phases: 253 cm$^{-1}$ of ZnSe [15-17], 251 cm$^{-1}$ of $Cu_2SnSe_3$ [8], and ~250 cm$^{-1}$ of amorphous selenium (a-Se) [18]. It is possible that selenium was present in CZTSe films, since they were fabricated in a Se-rich condition. a-Se tended to have a weak feature at ~235 cm$^{-1}$, due to the presence of crystalline phase trigonal Se (t-Se) [18]. Raman line near 303.6 cm$^{-1}$ from the finished device S3 is known to be from CdS [19], which is the window layer of the CZTSe solar device. The origin of the broad band ranging from 380 to 390 cm$^{-1}$ is unclear, but common to the three samples. It could be the second order feature of the ~196 cm$^{-1}$ mode. The purpose of this study is not trying to determine the exact origins of all these Raman features using Raman spectroscopy alone, but to demonstrate a methodology that can offer more information than the conventional Raman probe. Such information may contribute to the ultimate understanding of such a complex material system, when the results are correlative with those from other probing techniques.

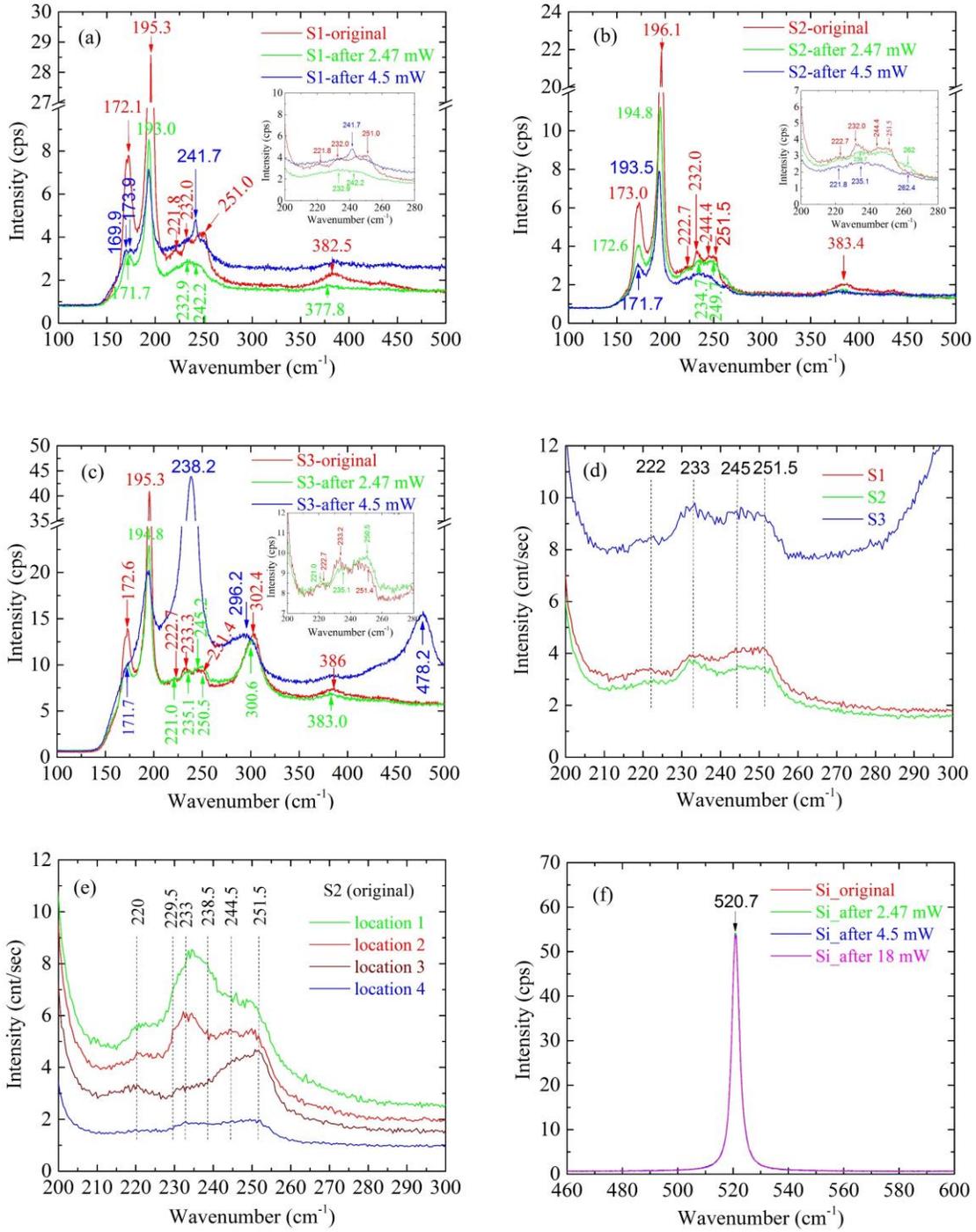

FIG. 2. Raman spectra of CZTSe and Si samples at 0.146 mW before and after being illuminated by high powers. (a) S1; (b) S2; (c) S3; (d) comparison of S1, S2, and S3; (e) comparison of different locations of S2; (f) Si.

A previous report indicated that Raman spectra taken at the same spot of CZTSe absorber with different laser powers (0.2 mW, 0.5 mW, and 2.5 mW) were the same except for a change in background [20]. However, in this work, we have found that even at 2.47 mW significant changes already occurred, not only at the illumination site but also in the region adjacent to that, and more interestingly sample dependent; while at 4.5 mW more drastic and sample dependent changes occurred both at the illumination site and in the adjacent region. Here we focus on the changes at the illumination site, and will discuss the changes of the adjacent region later when discussing the Raman mapping data.

After the 2.47 mW illumination, as shown by the spectra in Figs. 2(a) – (c) (in green), the two main CZTSe peaks at 195 – 196 cm$^{-1}$ and 171 – 172 cm$^{-1}$ experienced not only intensity reduction, but also red shifted in all three samples. For example, in S1, the CZTSe main peak at 195.3 cm$^{-1}$ shifts to 193.0 cm$^{-1}$, 172.1 cm$^{-1}$ to 171.7 cm$^{-1}$. Again the redshift of the 195-196 cm$^{-1}$ is the largest for S1, 2.3 cm$^{-1}$, compared to 1.3 cm$^{-1}$ for S2, and 0.5 cm$^{-1}$ for S3. These changes seem to suggest that the atomic bonds at the illuminated site were thermally expanded irreversibly as a result of the local heating. The weak features in the mesa-like band appear to smear out and weaken in both S1 and S2, but in S2 a new feature at ~262 cm$^{-1}$ emerges, as shown in Figs. 2(a) and (b); whereas in S3, those weak features remained but their relative intensities changed, for instance, the peaks at ~235 and 250.5 cm$^{-1}$ became more apparent, as shown in Fig. 2(c).

Table I. Summary of Raman modes from three CZTSe samples before and after HP illumination.

| Sample | before | after 1$^{st}$ illumination | after 2nd illumination | origin |
|---|---|---|---|---|
| S1 | 172.1 cm$^{-1}$ | 171.7 cm$^{-1}$ | 169.9 cm$^{-1}$ | CZTSe |
| | 195.3 cm$^{-1}$ | 193.0 cm$^{-1}$ | 193.0 cm$^{-1}$ | CZTSe |
| | 222 cm$^{-1}$ | | | CZTSe |
| | 233 cm$^{-1}$ | | | CZTSe |
| | 245 cm$^{-1}$ | | | CZTSe |
| | 251.5 cm$^{-1}$ | | | a-Se |
| | | | 241.7 | MoSe$_2$ |
| S2 | 173.0 cm$^{-1}$ | 172.6 cm$^{-1}$ | 171.7 cm$^{-1}$ | CZTSe |
| | 196.1 cm$^{-1}$ | 194.8 cm$^{-1}$ | 193.5 cm$^{-1}$ | CZTSe |
| | 222 cm$^{-1}$ | | | CZTSe |
| | 233 cm$^{-1}$ | | | CZTSe |
| | 245 cm$^{-1}$ | | | CZTSe |
| | 251.5 cm$^{-1}$ | | | a-Se |
| | | 262 cm$^{-1}$ | 262.4 cm$^{-1}$ | Cu$_x$Se$_y$ |
| S3 | 172.6 cm$^{-1}$ | 171.7 cm$^{-1}$ | ~172 cm$^{-1}$ | CZTSe |
| | 195.3 cm$^{-1}$ | 194.8 cm$^{-1}$ | 194.8 cm$^{-1}$ | CZTSe |
| | 222 cm$^{-1}$ | 222 cm$^{-1}$ | | CZTSe |
| | 233 cm$^{-1}$ | 233 cm$^{-1}$ | | CZTSe |
| | 245 cm$^{-1}$ | 245 cm$^{-1}$ | | CZTSe |
| | 251.1 cm$^{-1}$ | 250.5 cm$^{-1}$ | | a-Se |
| | | 235.1 cm$^{-1}$ | 238.2 cm$^{-1}$ | t-Se |
| | | | 478.2 cm$^{-1}$, 714 cm$^{-1}$ | t-Se |
| | 302.4 cm$^{-1}$ | 300.6 cm$^{-1}$ | 296.2 cm$^{-1}$ | CdS |

After illuminated further at 4.5 mW, with the spectra given in Figs. 2(a) – (c) (in blue), the intensity of the strongest CZTSe peak initially at 195 - 196 cm$^{-1}$ decreased further for all three samples, but the peak position remains nearly the same as that after the first HP illumination, except for in S2 it shifted further from 194.8 cm$^{-1}$ to 193.5 cm$^{-1}$, resulting in the largest redshift of 2.6 cm$^{-1}$ among all. For S1, as shown in Fig. 2 (a), the intensity of 193 cm$^{-1}$ peak is reduced to only 17% of its original value before any high power illumination, and the 172 cm$^{-1}$ mode becomes a weak shoulder from 168 cm$^{-1}$ to 178 cm$^{-1}$. Interestingly, for S2, the main CZTSe peak intensity only decreases by a factor of four, and the ~ 172 cm$^{-1}$ CZTSe peak remains sharp. For the spectral region of the mesa-like band, a sharp peak emerges at 241.7 cm$^{-1}$ as the most prominent feature in S1, but not in S2 and S3; no significant further change in S2 except for that the ~262 cm$^{-1}$ feature is slightly better resolved; in S3, one very strong sharp peak at 238.2 cm$^{-1}$ appeared, accompanying by a second order peak at 478.2 cm$^{-1}$ (and a third order peak at ~714 cm$^{-1}$, not shown), and a broad band at ~296 cm$^{-1}$. Obviously, the second HP illumination brought more qualitatively different changes to these samples. The contrast between S1 and S2 suggests that the CZTSe structure prepared by co-evaporation seems to be more robust than that prepared by sputtering against HP illumination induced structural modifications.

Regarding the ~242 cm$^{-1}$ peak in S1, although it was previously reported as a CZTSe Raman mode [21, 22], it actually matches the $A_{1g}$ Raman mode of $MoSe_2$ [23]. Since it only appeared after the HP illumination, at least it cannot be intrinsic to CZTSe. The weak ~262 cm$^{-1}$ peak in S2 turns out to be the Se-Se stretching mode of $Se_2$ ions in $Cu_xSe_y$ [24-27] coming from the periphery of the illuminated spot, which will be apparent in the surface Raman mapping result to be discussed later. The ~238 cm$^{-1}$ peak in S3 is most likely the $A_1$ mode of t-Se [18, 28]. It was shown that as a result of photo-crystallization, a broad band at ~ 250 cm$^{-1}$ and a shoulder at ~ 235

cm$^{-1}$ observed in an a-Se film became two resolved E symmetry mode at 233 cm$^{-1}$ and A$_1$ symmetry mode at 237 cm$^{-1}$ of t-Se [18]. This transformation matches our observations in S3 before and after the second HP illumination, which suggests the possibility that a-Se existed in the as-deposited CZTSe film; and after the second HP illumination, the a-Se was nearly completely crystallized to t-Se, since no ~ 250 cm$^{-1}$ peak remained. One may notice that the intensity at ~250 cm$^{-1}$ originally observed in the as-grown CZTSe film was much lower than the intensity of ~238 cm$^{-1}$ that appeared after the second HP illumination. It is possible that the partial decomposition of CZTSe provided more selenium, which contributed to the strong t-Se Raman mode. Similar changes caused by HP illumination were also observed from other similar co-evaporated CZTSe devices. Table I summarizes the Raman modes before and after the first and second high power illumination.

In summary, changes in Raman spectra from all three CZTSe samples after the HP illumination indicated partial decomposition and plastic changes of the materials. However, different samples responded rather differently to the HP illumination, indicating that seemingly similar CZTSe materials could differ significantly in their microscopic structures. As a comparison to a conventional semiconductor, Fig. 2(f) gives Raman spectra collected from a Si sample before and after HP illumination. Illuminated with different laser powers up to a maximum of 18.9 mW (4.5 x 10$^6$ W/cm$^2$) for 50 s to 400 s, the Si Raman spectra look almost the same as its original spectrum collected with 0.149 mW. It was also found previously that the highest power was in general safe for an epitaxial GaAs sample, but could induce a structure change when a dislocation type defect was illuminated [29]. Apparently, quaternary CZTSe is structurally not as robust as an elemental or binary semiconductor such as Si and GaAs. The contrast is mostly because of the difference in chemical bonding strength, but maybe also the structural defects and lower thermal conductivity of the polycrystalline film.

## 2. Raman mapping near the illuminated site

2D Raman mapping was performed to examine the spatial extension of the local heating effect caused by the HP illumination on the CZTSe samples, as shown in Figs. 3 – 5 for the 2.47 mW and Figs. 7 – 9 for the 4.5 mW illumination. One random spot from each sample surface was first illuminated by 2.47 mW for 100 seconds, Raman mapping was then acquired with 0.146 mW laser power from a 5 × 5 µm (in Fig. 3) or 8 × 8 µm (in Figs. 4 – 5 and 7 – 9) square centered at the illuminated spot. The optical images in Figs. 3 – 5 (a) show color changes to all three illuminated spots after the 2.47 mW laser illumination, especially one can see a halo surrounding the illuminated spot from S2. For each sample, Raman intensity mapping reveals a dark circle for the 196 cm$^{-1}$ peak, as shown in Figs. 3(b) – 5 (b), confirming the intensity reduction of the CZTSe main Raman mode at 196 cm$^{-1}$. In the Raman mapping for S1, a Raman peak at ~261 cm$^{-1}$ was observed in a ring outside of the illumination site, as evident in Fig. 3(c), whereas in the mapping results of S2 and S3, the ~261 cm$^{-1}$ peak was not observed, and the intensity maps of 220 – 270 cm$^{-1}$ showed no significant spatial variation, as shown in Fig. 4(c) and Fig. 5(c). The appearance of the ~261 cm$^{-1}$ peak suggests the formation of Cu$_x$Se$_y$ [24-27], occurring mostly at some distance away from the illumination site, which was likely resulted from the specific temperature gradient caused by the laser heating. The formation of the ring structure also explains why when measured at the illumination site, the Cu$_x$Se$_y$ feature was not resolved in S1 and S3, very weak in S2, as shown in Fig. 2(b). In Fig. 6(a), a comparison is given between the spectra measured at the illumination site and a typical spot on the Cu$_x$Se$_y$ ring, which shows that a sharp 261.1 cm$^{-1}$ peak appeared from the Cu$_x$Se$_y$ ring but not the illumination site. Additionally, on the Cu$_x$Se$_y$ ring, the CZTSe modes 195.7 and 171.7 cm$^{-1}$ show no significant shift, and the intensity of the CZTSe 195.7 cm$^{-1}$ mode is significantly stronger than that of the illumination site, but only half of the original value from a general S1 spot before any HP illumination, indicating that HP illumination might have led to partial CZTSe decomposition on the Cu$_x$Se$_y$ ring, though not as much as the illumination site.

The Raman mapping of the primary CZTSe mode at the 196 cm$^{-1}$ reveals most directly the spatial extension the material being affected by the local heating. For S1, as shown in Fig. 3(a), beyond the darkest ~1 µm circle (comparable to the laser spot size of ~0.76 µm) at the illumination site, a dark area with a diameter of ~3 µm, comparable to the size of the ~261 cm$^{-1}$ ring, also showed significantly lower intensity of the 196 cm$^{-1}$ peak. On the other hand, 2.47 mW laser illumination only affected an area of about 1.5 µm diameter in S2 and even less than 1 µm diameter in S3. The difference between S1 and S2 seems to suggest that the two nominally similar films might have rather different thermal conductivities, which might be partially responsible to the different responses in structural change mentioned above. The extra layers above CZTSe in the device S3 might improve the thermal conductivity of the structure as a whole, thus, showing the least spatial extension. The contrast between S1 and S2 suggests that the CZTSe absorber layer prepared by the sputtering method is more sensitive to the HP illumination than the film fabricated by the co-evaporation method.

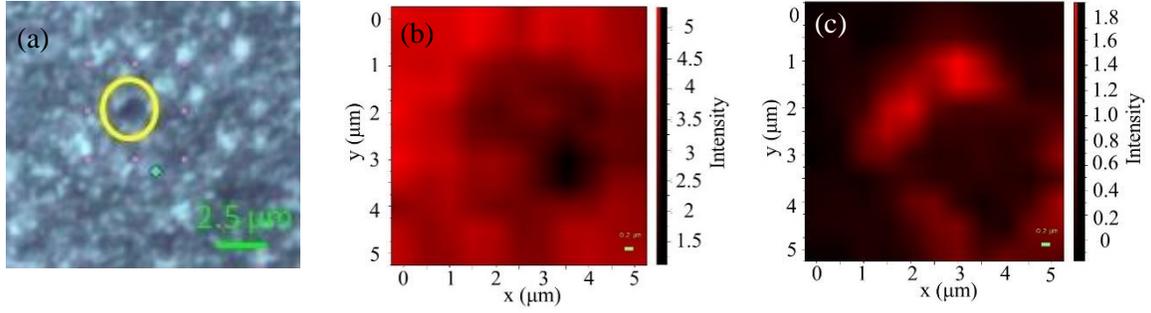

FIG. 3. S1 Raman mapping after one spot being illuminated by 2.47 mW for 100 s: (a) optical image of the illuminated spot; (b) Raman mapping of 196 cm$^{-1}$ peak; (c) Raman mapping of 261 cm$^{-1}$ peak.

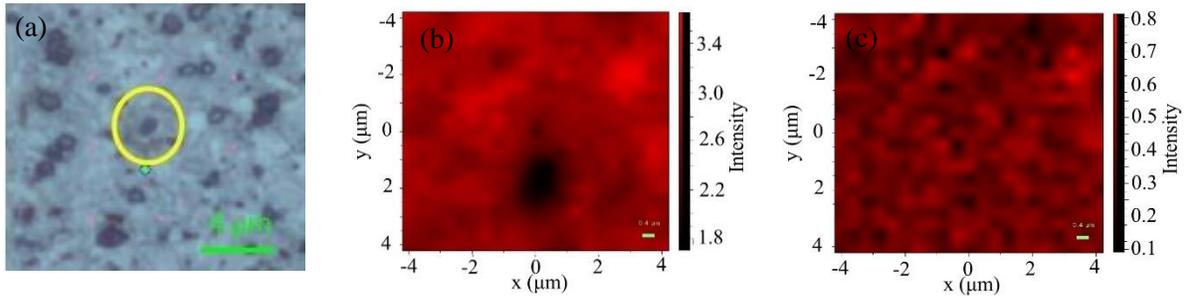

FIG. 4. S2 Raman mapping after one spot being illuminated by 2.47 mW for 100 s: (a) optical image of the illuminated spot; (b) Raman mapping of 196 cm$^{-1}$ peak; (c) Raman mapping of 220 ∼ 270 cm$^{-1}$.

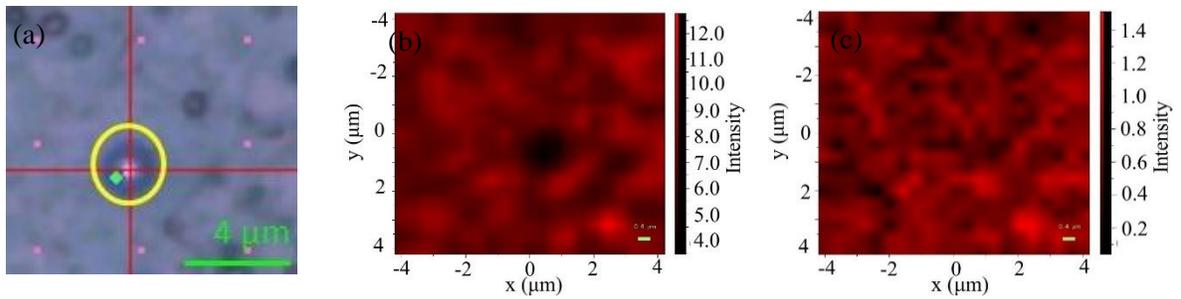

FIG. 5. S3 Raman mapping after one spot being illuminated by 2.47 mW for 100 s: (a) optical image of illuminated spot; (b) Raman mapping of 196 cm$^{-1}$ peak; (c) Raman mapping of 220 ∼ 270 cm$^{-1}$.

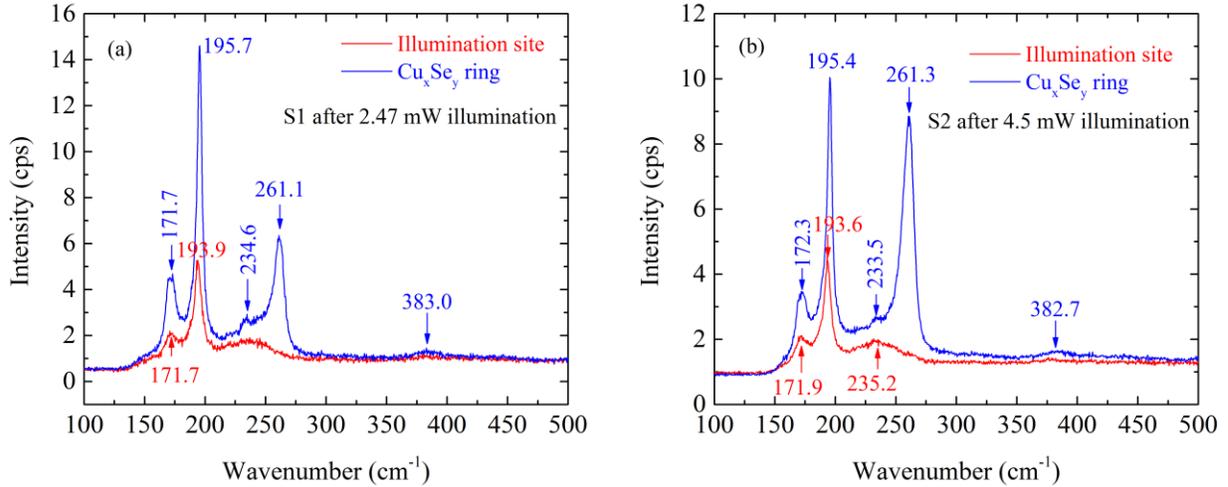

FIG. 6. Comparison of Raman spectra measured at illumination site and $Cu_xSe_y$ ring: (a) S1 after 2.47 mW illumination; (b) S2 after 4.5 mW illumination.

Figs. 7 – 9 show the mapping results with 0.146 mW after the 4.5 mW illumination. Larger affected regions were revealed in optical images in Figs. 7 (a) – 9 (a). Raman mapping in Fig. 7 (b) showed that the 196 cm$^{-1}$ peak intensity of S1 reduced in a circular region with a diameter larger than 6 µm. The affected area of S2 also enlarged to about 5 µm in diameter. For S3, the affected area only has a diameter about 2 µm. The extra layers in the CZTSe device again offered some protection to the CZTSe absorber. In the intensity map of the 261 cm$^{-1}$ peak, a $Cu_xSe_y$ rich ring still existed in S1 as shown in Fig. 7 (c), but now a similar $Cu_xSe_y$ ring was also appeared, although smaller in S2, as shown in Fig. 8(c). Fig. 6 (b) compares the representative spectra measured at the ring and the illuminated site. Similar with S1, strong $Cu_xSe_y$ Raman signal was only observed from the ring, and the CZTSe main peak showed stronger peak intensity and smaller redshift compared with the illumination site. The size of the $Cu_xSe_y$ rich ring reflects that of the heated region reaching the phase transition temperature, which in turn depends on the thermal conductivity of the material. For S3, consistent with the observation of a strong Raman peak at ~238 cm$^{-1}$, Fig. 9 (c) shows a map of 238 cm$^{-1}$, showing a bright region that is comparable in size of the dark region of Fig. 9(b).

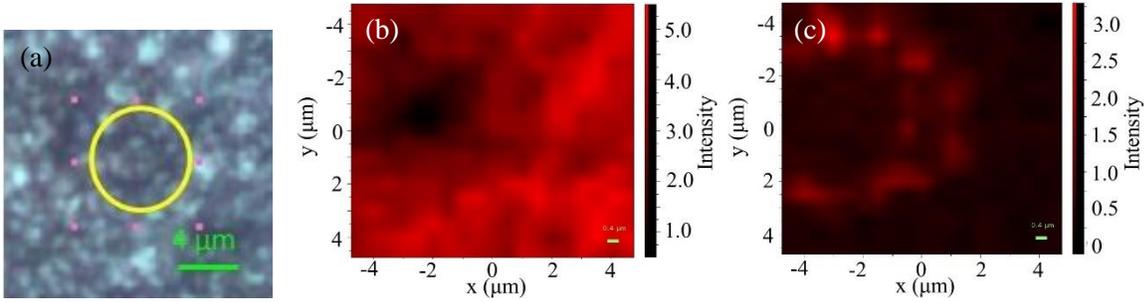

FIG. 7: S1 Raman mapping after one spot being illuminated by 4.5 mW for 36 s: (a) optical image of illuminated spot; (b) Raman mapping of 196 cm$^{-1}$ peak; (c) Raman mapping of 261 cm$^{-1}$ peak.

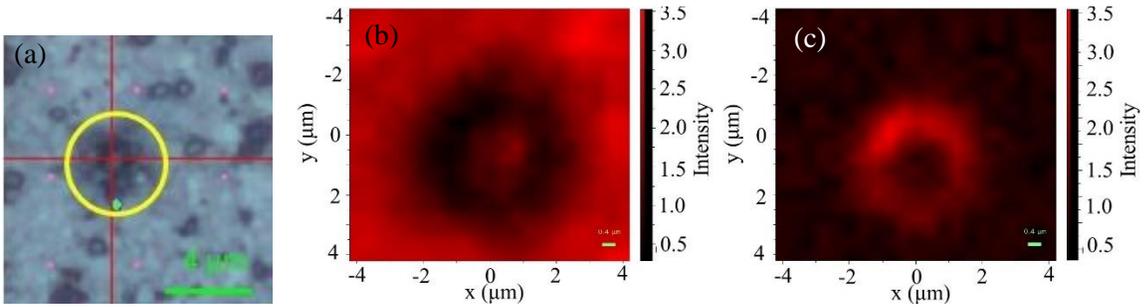

FIG. 8. S2 Raman mapping after one spot being illuminated by 4.5 mW for 36 s: (a) optical image of illuminated spot; (b) Raman mapping of 196 cm$^{-1}$ peak; (c) Raman mapping of 261 cm$^{-1}$ peak.

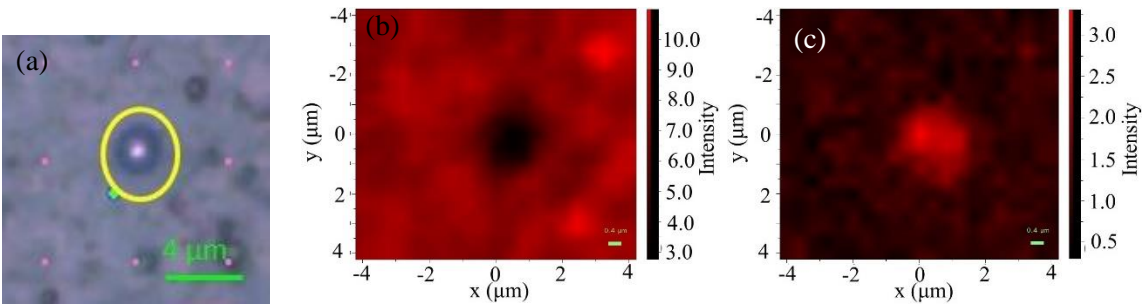

FIG. 9. S3 Raman mapping after one spot being illuminated by 4.5 mW for 36 s: (a) optical image of illuminated spot; (b) Raman mapping of 196 cm$^{-1}$ peak; (c) Raman mapping of 238 cm$^{-1}$ peak.

## B. High temperature studies of CZTSe samples

High temperature Raman spectra of a random spot from each sample were collected with a step of 20 °C starting from room temperature using 0.147 mW laser power with a 50x long working distance lens. Raman spectra of S1 at representative temperatures were shown in Fig. 10 (a). The main CZTSe peak at ~196 cm$^{-1}$ started red-shifting when temperature raised to above 40 °C, at a rate about 0.0212 ±0.0007 cm$^{-1}$/°C, which is obtained from the linear fit in Fig. 10(b). As indicated in Fig. 1, at this power level, a small heating is expected for this sample. The residual laser heating could explain why no shift was observed when the sample temperature was below 40 °C, as shown in Fig. 10(b). Using the obtained slope, we could estimate the laser induced local heating found in Fig. 1 to be around 20 °C, and the main CZTSe Raman mode at 0 °C should be about 196.8 cm$^{-1}$ according to the linear fit. When temperature was raised up to 460 °C, the primary CZTSe features almost disappeared, but a new weak peak at 821 cm$^{-1}$ was observed, which is close to a molybdenum oxide Raman mode at 819 cm$^{-1}$ [30]. Another CZTSe Raman mode at 172 cm$^{-1}$ also experienced redshift with raising temperature, but showing smaller slope, as shown in Fig. 10(b), and little change in intensity till reaching around 200 °C. At higher temperatures, it started to show significant intensity reduction and peak broadening and finally became a weak shoulder in the spectrum at 400 °C. The mesa-like band from 220 cm$^{-1}$ to 250 cm$^{-1}$ became narrower and forming a broad peak centered around 230 cm$^{-1}$ with increasing temperature. Additional random spots of this sample were also measured at 480 °C, with two typical types of Raman spectra shown in Fig. 10(c). If the result of the original tested spot is referred to as Type 1, Type 2 spectrum from some other locations turned out to be very different. In addition to the 185 cm$^{-1}$ and 220 – 250 cm$^{-1}$ band in Type 1, there were multiple extra weak peaks at ∼288 cm$^{-1}$, ∼661 cm$^{-1}$, ∼817 cm$^{-1}$, etc. in Type 2. By comparing the two Raman spectra in Fig. 10(c), one can find the anti-correlation between 184 cm$^{-1}$ and 821 cm$^{-1}$ peak. This is reasonable since strong 821 cm$^{-1}$ Raman line indicated the oxidization of Mo layer below the CZTSe layer. In other words, layers above molybdenum at these locations were largely gone.

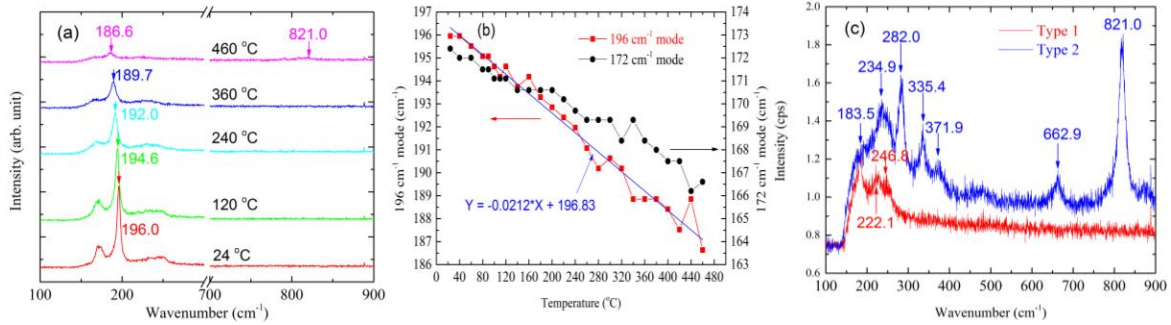

FIG. 10. Raman study of S1 measured at different temperatures up to 480 °C and CZTSe main peak shift with temperature: (a) Raman spectra at different temperatures; (b) CZTSe main peak shift with increasing temperature; (c) Two types of Raman spectrum measured from S1 surface at 480 °C.

The sample was then cooled down to room temperature and examined by using the 100x lens. As shown in Fig. 11(a), there were three typical types of regions on the sample surface (highlighted by colored circles): very bright (green), dark (red), and greyish (blue) which could be a transition state between the other two. Fig. 11(b) are Raman spectra from these three types of spots. The grey spot shows very strong Raman lines at 253 cm$^{-1}$ and 491 cm$^{-1}$ which match the LO and 2LO Raman modes of ZnSe [16]. From the black spot, besides the ZnSe peaks and $Cu_2SnSe_3$ mode at 179.2 cm$^{-1}$ [8], two additional peaks at 209.0 cm$^{-1}$ and 216.9 cm$^{-1}$ were observed. ZnSe and CuSe were both reported generating a Raman line near 209 cm$^{-1}$ [15, 31, 32]. Here, with the coexistence of a strong ZnSe Raman line at 253 cm$^{-1}$, this 209 cm$^{-1}$ is assigned as ZnSe TO Raman mode. On the bright spot, only very weak CZTSe peaks at 197 cm$^{-1}$ was found. The multiple Raman peaks from 245.2 cm$^{-1}$ to 819.2 cm$^{-1}$ observed from the bright spot are all very close to the reported orthorhombic $MoO_3$ Raman bands at 247, 284, 292, 339, 367, 380, 469, 667 and 822 cm$^{-1}$ [30]. The origins of other Raman peaks at 158.3 cm$^{-1}$, 858.8 cm$^{-1}$ and 882 cm$^{-1}$ from the bright spot are unclear. By examining the S1 sample after high temperature measurements, it can be concluded that almost the whole CZTSe absorber has decomposed after being heated to 480 °C. In most cases, reaction $Cu_2ZnSnSe_4 \rightarrow Cu_2SnSe_3 + ZnSe$ took place and the surface turned darker than its original color. There are also regions left with only ZnSe as the greyish areas. The highly reflective bright spot mainly comes from the Mo substrate which has been oxidized to $MoO_3$.

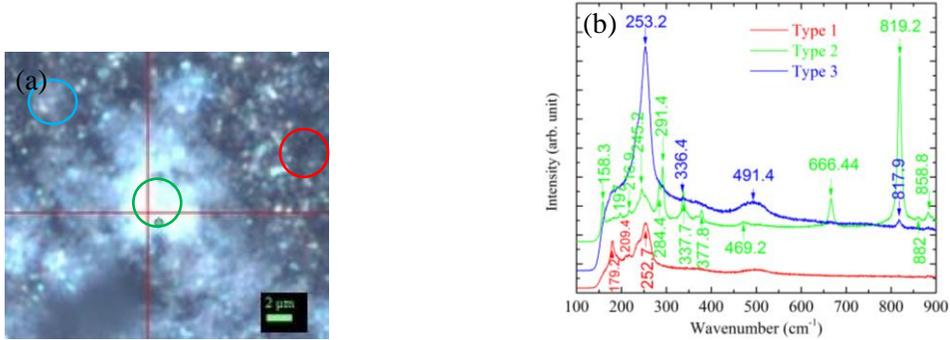

FIG. 11. Three different types of regions on S1 surface when back to room temperature: (a) optical image; (b) Raman spectra after high temperature measurements.

Another piece from S1 was heated up to 860 °C by increasing the temperature at a step of 40 °C. The spectra are shown in Fig. 12(a) for selected temperatures. When temperature was lower than 540 °C, the main changes in the CZTSe Raman mode such as intensity reduction and linewidth broadening were similar with the results of the first piece shown in Fig. 10(a). As temperature was further increased, the band from 220 $cm^{-1}$ to 250 $cm^{-1}$ started to become narrower and sharper, finally formed a sharp and strong peak centered at ~237 $cm^{-1}$ at 580 °C, together with the appearance of its second order mode at ~478 $cm^{-1}$. These two peaks reach its maximum intensity at around 700 °C and gradually diminish at higher temperatures. As described in the S3 high laser power illumination study, the strong peak at ~237 $cm^{-1}$ and ~478 $cm^{-1}$ are t-Se modes [18]. Interestingly, Fig. 12(b) shows that the main CZTSe peak reached the largest red shift in the temperature range of 500 ~ 600 °C. When the 238 $cm^{-1}$ peak started to emerge, i. e., when temperature was higher than 580 °C, the ~186 $cm^{-1}$ peak shifted to higher wavenumbers with increasing temperature. We note that the ~186 $cm^{-1}$ Raman peak observed at temperature above 500 °C might not be CZTSe related but likely from a secondary phase $Cu_2SnSe_3$ or ZnSe, since the CZTSe peak was not observed when the sample was brought back to room temperature, but with multiple new Raman peaks that did not exist before the heating process, as shown in Fig. 12(c). Among them, 346, 565 and 732 cm$^{-1}$ are from molybdenum oxide. The origins of the rest peaks are not clear. However, the CZTSe related Raman modes were totally gone.

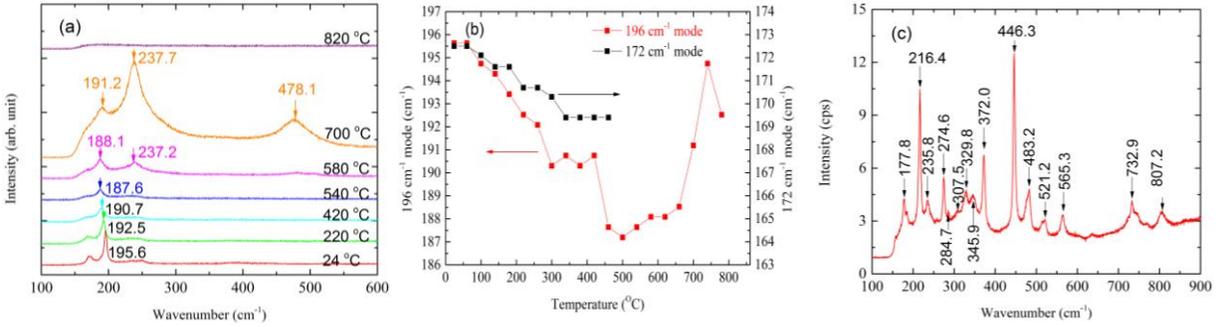

FIG. 12. A second high temperature study of S1 up to 860 ºC. (a) Raman spectra at different temperatures; (b) CZTSe main peak shift with increasing temperature; (c) Room temperature Raman spectrum after the high temperature study.

A piece of S2 was also studied at high temperatures up to 900 °C with a temperature step of 20 °C. Typical Raman spectra at selected temperatures are shown in Fig. 13(a), and the shift of the main CZTSe peak of ~196 cm$^{-1}$ with increasing temperature is plotted in Fig. 13(b). Similar with S1, t-Se gradually became the dominant component when temperature was higher than 600 °C, reaching its maximum intensity at around 700 °C. At the same time, the 187 cm$^{-1}$ peak was shifting to higher wavenumbers. No Raman signal could be detected at 900 °C. When back to room temperature, three representative types of Raman spectra were observed, as shown in Fig. 13(c). All peaks in Type 1 spectrum were from $MoO_2$ [30], whereas Type 3 spectrum composes of $Cu_2ZnSe_3$ at 180.8 cm$^{-1}$ [8] and ZnO at 438 cm$^{-1}$ [33, 34]. Type 2 spectrum contains both $MoO_2$ and ZnO Raman modes as well as another peak from $Mo_4O_{11}$ at ∼305 cm$^{-1}$ [30].

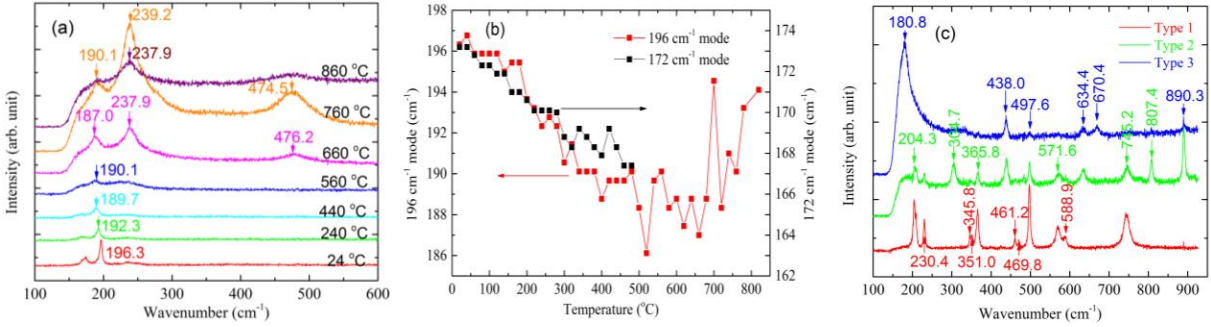

FIG. 13. Raman spectra of S2 measured at different temperatures and CZTSe main peak shift with temperature: (a) Raman spectra at different temperatures; (b) CZTSe main peak shift with increasing temperature; (c) Room temperature Raman spectra of three types of spots from S2 after the high temperature study.

Now, the t-Se modes have also been observed from both bare CZTSe film samples (S1 and S2) after being heated. We may conclude that ~238 and 478 cm$^{-1}$ peaks of Fig. 2 (c) for S3 after the HP illumination were not from the extra layers of the CZTSe device, since the bare film samples did not have the extra layers beyond the absorber.

CZTSe device S3 was also examined at high temperatures up to 800 °C with results shown in Fig. 14(a). From 380 °C to 660 °C, the t-Se modes were very weak. It only became clear and dominant when temperature was higher than 700 °C and disappeared at around 800 °C. Due to the surface roughness, broad linewidth, and low signal intensity at high temperatures, the positions of the main CZTSe peak in S3, as shown in Fig. 14(b), could not be as accurately determined as in S1 and S2, but in general agreement with them. Back to room temperature, two types of Raman spectra were found in S3, as shown in Fig. 14(c): one shows only one strong peak of SnSe$_2$ at 183 cm$^{-1}$; and the other shows Raman peaks from a mixture of MoO$_2$, Mo$_4$O$_{11}$, MoO$_3$ and ZnO.

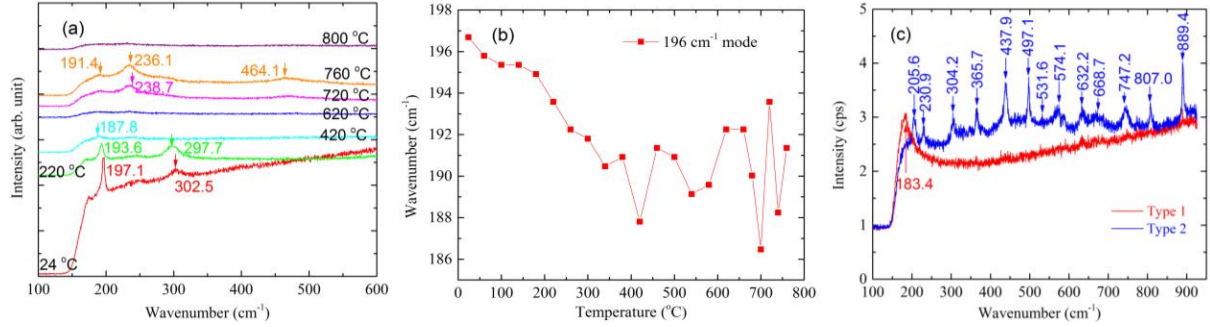

FIG. 14. Raman spectra of S3 at different temperatures and CZTSe main peak shift with temperature: (a) Raman spectra at different temperatures; (b) CZTSe main peak shift with increasing temperatures; (c) Room temperature Raman spectra of two types of spots of S3 after the high temperature study.

Another piece of S3 was also tested with temperature increased at a step of 20 °C up to 400 °C and then cooled down to room temperature. Most parts of the absorber were found to recover at least 60% of its original intensity for the ~196 cm$^{-1}$ CZTSe main peak, although the frequency has redshifted significantly to 192.6 cm$^{-1}$, whereas the CdS mode at ~ 303 cm$^{-1}$ could hardly be observed. Some dark spots with bright halo were observed. Fig. 15 shows four Raman spectra obtained from spots with different distance away from one such black spot, with its optical image shown as an inset in Fig. 15. At the center of the dark spot, resonant t-Se Raman modes were observed at 241, 482 and 717 cm$^{-1}$ together with a shifted CZTSe peak at 192.6 cm$^{-1}$. The farther away from the black spot center, the smaller is the Se Raman signals. The observations indicate that after the high temperature measurements, the dark center on this device surface became rich in selenium, and the decomposition process at high temperature within the film was not uniform. As suggested by the similarities between S2 and S3 in their high temperature tests, we predict that t-selenium could in principle also appear in S2, if it were illuminated by high enough laser power, although the material would be seriously ablated. Perhaps the processes to make the absorber material a finished device cause the absorber more sensitive to high laser power illumination. We also notice that the three orders of t-Se Raman mode shown in Fig. 15 induced by uniform heating are systematically blue-shifted from those in Fig. 2(c) by HP illumination, suggesting that they were in different states of strain.

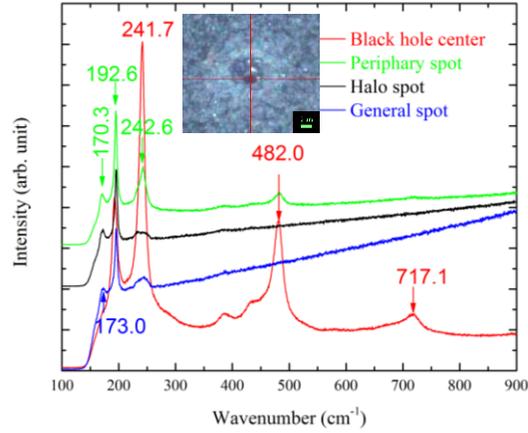

FIG. 15. Room temperature Raman spectra of different spots from S3 surface after the sample being heated to 400 °C.

$Cu_xSe_y$ as a new phase was often observed after HP illumination. However, no $Cu_xSe_y$ mode at 261 cm$^{-1}$ was observed in the high temperature measurements from any of the three samples. In the high temperature study, the entire sample was heated uniformly, whereas laser illumination offered highly localized non-uniform heating. Thus, lateral thermal conductivity and stress both might play a role in the formation of the $Cu_xSe_y$ peak.

Finally, the local temperature of the illuminated site induced by HP illumination was estimated by comparing the Raman shift of the main CZTSe Raman peak between the lower and high power spectrum. Based on the red-shifts observed with the 2.47 mW illumination: 4.9 (S1), 3.6 (S2), and 2.2 (S3) cm$^{-1}$, and the temperature dependences shown in Figs. 10, 13, 14, we estimated that the local temperature reached about ~280 °C in S1, ~260 °C in S2, and ~200 °C in S3. The results of the higher power 4.5 mW illumination are inappropriate for such estimate, because the CZTSe material right at the illumination site was either ablated or converted into t-Se (see Fig. 9), thus, the measured CZTSe signal of much reduced intensity was actually from the adjacent area that might not be heated as much. Furthermore, HP illumination might induce inelastic local strain, which could make the local heating effect more complicated than a simple temperature change. More dedicated studies are required to carefully probe the evolution from elastic to plastic change under local heating.

## C. EDS after HP illumination

EDS was also performed to study the elemental compositions near the illuminated site after HP illumination. The SEM images of 4.5 mW illuminated sites from S1 and S2 are shown in Fig. 16, and the elemental ratios are summarized in Table II. After the HP illumination, the illuminated center of S1 showed the highest Se concentration, but lowest Cu concentration, which is consistent with the formation of $MoSe_2$ at the illuminated site. Oppositely, the S2 illuminated center showed the lowest Se and Cu concentration but highest Zn and Sn concentration, indicating different structure changes after HP illumination in these two films. For both S1 and S2, the illuminated sites have very low Cu/(Zn+Sn) ratios, however, the underlying changes are somewhat different. Away from the illuminated site, the bright ring in Fig. 16(b) in S2 exhibited higher Cu and Se concentrations as well as a higher Cu/(Zn+Sn) ratio than the general area, which is consistent with the Raman mapping result that a $Cu_xSe_y$ rich ring surrounding the illuminated center appeared after 4.5 mW illumination. However, Raman mapping also suggests a $Cu_xSe_y$ ring on the S1 surface, but the $Cu_xSe_y$ rich region did not appear bright in the SEM image of Fig. 16(a). Also, by comparing the size of dark illumination center, one can find that S1 is much larger than S2, which again confirms the surface Raman mapping results in Figs. 3 and 4 and 7 and 8.

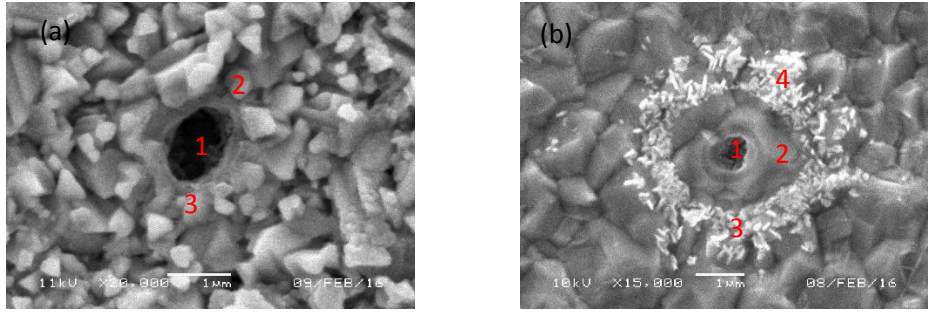

FIG. 16. SEM images of S1 and S2 surface spots being illuminated by 4.5 mW: (a) S1; (b) S2.

Table II. EDS results of S1 and S2 after surface spots being illuminated by 4.5 mW. The numbers in the parenthesis are the nominal values for the samples.

|    | Regions | Cu (%) | Zn (%) | Sn (%) | Se (%) | Zn/Sn | Cu/(Zn+Sn) |
|---|---|---|---|---|---|---|---|
| S1 | 1 | 7.3 | 12.6 | 9.3 | 70.8 | 1.35 | 0.33 |
|    | 2 | 17.5 | 16.9 | 11.1 | 54.5 | 1.52 | 0.67 |
|    | 3 | 19.6 | 16.1 | 9.3 | 55.0 | 1.73 | 0.71 |
|    | General area | 18.9 | 15.6 | 9.1 | 56.4 | 1.71 (1.25) | 0.76 (0.85) |
| S2 | 1 | 16.3 | 31.0 | 19.1 | 33.6 | 1.62 | 0.32 |
|    | 2 | 17.8 | 16.9 | 13.4 | 51.9 | 1.26 | 0.59 |
|    | 3 | 22.9 | 11.8 | 9.9 | 55.4 | 1.19 | 1.05 |
|    | 4 | 21.8 | 12.4 | 9.9 | 55.9 | 1.25 | 0.99 |
|    | General area | 21.1 | 12.6 | 10.0 | 56.3 | 1.26 (1.32) | 0.94 (0.82) |

## IV. Summary

In this work, combining spatially-resolved high-power (HP) and high-temperature (HT) Raman spectroscopy, we have revealed the significant difference in microscopic structures of CZTSe films which would otherwise appear very similar using conventional characterization tools, such as Raman and XRD. HP illumination brought different changes to films prepared by different method. At the illumination site, CZTSe Raman modes exhibited different amounts of peak shift, indicating different degrees of plastic change after HP illumination. In general, the sputtered film is structurally less robust than the co-evaporated films. Raman mapping shows the different spatial extension of the impact of local HP illumination, suggesting different thermal conductivities of the samples fabricated by different methods. For the two bare CZTSe films, a $Cu_xSe_y$ rich ring enclosing the illuminated site was induced by HP illumination, but the size of the ring was found to be significantly larger for the sputtered sample than the co-evaporated one, and also a new phase possibly $MoSe_2$ was formed at the illumination site for the former but not the latter. In the CZTSe device, the appearance of strong Raman modes of ~238, ~478, and ~714 cm$^{-1}$ at the illumination site indicated photo-crystallization from a-Se to t-Se after HP illumination. HT Raman measurements on one hand reveals the different decomposition processes of the three different samples, and on the other hand, further confirms the t-Se originating from photo-crystallized a-Se or CZTSe decomposition. In the HPHT Raman measurements, the different responses of CZTSe films suggest the differences in microscopic structure, despit the macroscopically similar films. The combination of high power and high temperature Raman study can offer an effective approach to examine the microscopic structural variations of the complex alloys like CZTSe.

**Acknowledgement**

We are very grateful to Dr. I. Repins (NREL) for providing the co-evaporated samples and helpful discussions. Y.Z. acknowledges the support of Bissell Distinguished Professorship.